\documentclass[prb,twocolumn,showpacs,preprintnumbers,amsmath,amssymb]{revtex4}

\usepackage{epsfig}
\usepackage{graphicx}% Include figure files
\usepackage{dcolumn}% Align table columns on decimal point
\usepackage{bm}

\newcommand{\e}{{\bf e}}

\newcommand{\q}{{\bf q}}
\renewcommand{\r}{{\bf r}}

\newcommand{\I}{{\rm i}}

\def\gl{\lower.35em\hbox{$\stackrel{\textstyle>}{\textstyle<}$}} 
\def\gapp{\lower.35em\hbox{$\stackrel{\textstyle>}{\textstyle\sim}$}} 
\def\lapp{\lower.35em\hbox{$\stackrel{\textstyle<}{\textstyle\sim}$}} 
 
\begin{document}
\title{Plasmons and near-field amplification in double-layer graphene}
\author{T. Stauber and G. G\'omez-Santos}
\affiliation{Departamento de F\'{\i}sica de la Materia Condensada  and Instituto Nicol\'as Cabrera, Universidad Aut\'onoma de Madrid, E-28049 Madrid, Spain}
\date{\today}
\begin{abstract}
We study the optical properties of double-layer graphene for linearly polarized evanescent modes and discuss the in-phase and out-of-phase plasmon modes for both, longitudinal and transverse polarization. We find a energy for which reflection is zero, leading to exponentially amplified transmitted modes similar to what happens in left-handed materials. For layers with equal densities $n=10^{12}$cm$^{-2}$, we find a typical layer separation of $d\approx500\mu$m to detect this amplification for transverse polarization which may serve as an indirect observation of transverse plasmons. When the two graphene layers lie on different chemical potentials, the exponential amplification either follows the in-phase or out-of-phase plasmon mode depending on the order of the low- and high-density layer. This opens up the possibility of a tunable near-field amplifier or switch.
\end{abstract}
\pacs{73.21.Ac, 42.25.Bs, 78.67.Wj, 73.20.Mf}
\maketitle
%%%%%%%%%%%%%%%%%%%%%%%%%%%%%%%%%%%%%%%%%%%%%%%%%%%%%%%%%%%%%%%%%%%
% Section : Introduction
%%%%%%%%%%%%%%%%%%%%%%%%%%%
\section{Introduction}
The optical properties of graphene have attracted much attention recently.\cite{Geim09} A hallmark is given by the universal absorption of 2.3\% in the visible regime independent of the incident wavelength.\cite{Kuzmenko08,Nair08} For infrared frequencies, the absorption can be altered by varying the Fermi energy via an applied gate voltage\cite{Novoselov04}. For fixed chemical potential, the optical conductivity is then characterized by three regimes: i) the Drude peak related to intraband transitions, ii) a region with suppressed absorption due to Pauli-blocking and iii) the nearly constant plateau with $\sigma_0=\frac{\pi}{2}\frac{e^2}{h}$ due to interband transitions.\cite{Li08,Peres08}

There is also high potential to use graphene in information and communication technologies. Apart from applications as conductive and optically transparent electrode in resistive touchscreens of transparent and flexible displays,\cite{Bae10,Bonaccorso10} there are various examples to use graphene for opto-electronic devices such as photodetector\cite{Mueller10}, broadband absorber\cite{Liu11}, or mode-locked laser due to current saturation\cite{Sun10}. There are also interesting theoretical proposals not yet observed like the dynamical generation of a gap\cite{Fistul07} or carrier multiplication due to strong electron-electron interaction\cite{Song11}.

Here, we analyze the optical properties of double-layer graphene which is of potential interest due to the recent advances in fabricating graphene-boron nitride structures.\cite{Dean10,Geim11} Our main result is that there is exponential amplification of evanescent modes in gated double-layer systems of graphene, first discussed by Pendry in the context of so-called left-handed materials with negative refraction index.\cite{Pendry00} The appearance of exponential amplification in between two layers is closely related to the existence of surface plasmon polaritons as noted by Pendry and later discussed by Haldane\cite{Haldane02}. This effect can be considered the evanescent equivalent of the extraordinary transmission observed in perforated metal sheets, where plasmons are believed to be involved.\cite{Ebbesen98,Moreno01,Vidal10}

In graphene, there are two types of (surface) plasmons, longitudinal (or transverse magnetic)\cite{Wunsch06,Hwang07} and transverse (or transverse electric) modes\cite{Ziegler07} which give rise to different behavior. Theoretically, these low-energy collective excitations have attracted much interest\cite{Polini08,Roldan09,GomezSantos09,SSP10,Tudorovskiy10,Hwang10,Gamayun11,Pedersen11} and recently, it was shown that they can be excited via a attenuated total reflection structure,\cite{Bludov10} paving the way to graphene plasmonics.\cite{Ju11,Koppens11,Moreno11} Experimentally, they have been observed via electron energy loss spectroscopy\cite{Seyller08,Tegenkamp11} and near-field nanoscopy.\cite{Basov11} 

Here, we will discuss how the plasmon modes are modified in a double-layer structure for both polarizations. In the case of longitudinal polarization, the two plasmon modes will split, if the layer separation is small enough such that the plasmon modes can still considerably interact electrostatically. This leads to an ordinary two-dimensional plasmon with $\sqrt{q}$ dispersion but larger energy where the charges of the two layers oscillate in phase, and a linear (acoustic) plasmon mode where the charges oscillate out of phase. Their dependence on the Fermi energy of the two layers has first been discussed in Ref. \cite{Hwang09} for $T=0$, for finite temperature in Ref. \cite{Vazifehshenas10}, and the plasmon interaction and hybridization in pairs of neighboring nanoribbons in Ref. \cite{Christensen11}. Here, we also discuss the scattering problem. In the case of transverse polarization, plasmons are not formed via charge accumulation and are only weakly bound to the graphene layer. These collective excitations are linked to the two-band structure of the electronic carriers and thus also exist in bilayer graphene where they become more localized.\cite{Jablan11} In a double layer structure, we find either one mode (for small layer separation) or two plasmon modes (for large layer separation).

As noted before, in the case of two plasmon modes, there exists a frequency $\omega_{exp}$ where reflection is zero and the transmission is exponentially amplified. For equal electronic densities in the two layers, this frequency lies between the two plasmon frequencies and is pinned to the out-of-phase mode for small layer separation. For large layer separation, the two plasmon modes merge and symmetrically sandwich the frequency of exponential amplification just as in the original model proposed by Pendry.\cite{Pendry00} For different electronic densities, $\omega_{exp}$ can follow either the low- or high-density plasmon mode depending on the order of the layers. 

We also discuss the amplification $\mathcal{A}$ proportional to the ratio of the intensity at the second interface to one at the first interface. The maximum of $\mathcal{A}$ lies between the two plasmon frequencies but only coincides with $\omega_{exp}$ in the case of large layer separation. This tunable plasmonic evanescent amplification could be useful for near-field microscopies.

The paper is organized as follows. In section II, we present the scattering problem and discuss the general behavior. In section III, we discuss the case of equal electronic density in the two layers and in section IV, we discuss the density-imbalanced system. We close with a summary and conclusions. In an appendix, the explicit formulas for the general setup with different dielectric media are given. To model graphene, we will only depend on the current-current correlation function at zero temperature which was evaluated in Ref. \cite{Principi09} within the Dirac cone approximation and in Ref. \cite{Stauber10} for the hexagonal lattice.

%%%%%%%%%%%%%%%%%%%%%%%%%%%%%%%%%%%%%%%%%%%%%%%%%%%%%%%%%%%%%%%%%%%
% Section : Scattering problem
%%%%%%%%%%%%%%%%%%%%%%%%%%%
\section{Scattering problem} To discuss the scattering behavior of linearly polarized light in a layered geometry, a convenient gauge condition is to choose the scalar potential to be zero. We thus only have to consider the vector potential. 

In the following, we will assume the graphene layers to be parallel to the $xy$-plane. For transverse polarized light with say $\q=q \e_x$, only the $y$-component $A_y$ is non-zero. Since here we are interested in the scattering behavior of evanescent modes, we define the {\em real} transverse wave number $q_i^\prime=\sqrt{q^2-\epsilon_i(\omega/c)^2}$, where $q$ denotes the conserved wavenumber parallel to the plane and $\epsilon_i$ the dielectric constant of the corresponding medium. For two dielectric media separated by graphene located at $z=0$, the vector potential of an incoming field reads
\begin{align}
\label{AnsatzA}
A_y(\r,z)=\sum_\q e^{\I \q\cdot\r}\left\{
\begin{array}{ll}
e^{-q_1'z}+Re^{q_1'z}&,z<0\\
Te^{-q_2'z}&,z>0
\end{array}
\right. 
\end{align}
A similar expression holds for longitudinal polarization,\cite{Falkovsky07} see also the appendix. 

For transverse polarization, the vector potential is continuous at the interface and the first derivative makes a jump due to the current generated by the vector field inside the graphene plane. A similar situation occurs for longitudinal polarization (see appendix). The current is related to the corresponding current-current susceptibility $\chi^{\parallel,\perp}$,\cite{Principi09,Stauber10} where $\parallel$ stands for longitudinal or $p$ and $\perp$ for transverse or $s$ polarization. This results in the following expressions for the transmission amplitude:
\begin{align}
T^\perp&=\frac{2q_1^\prime}{q_1^\prime+q_2^\prime+\frac{\chi^\perp(\q,\omega)}{\varepsilon_0c^2}}\\
T^\parallel&=\frac{2\epsilon_1q_2^\prime}{\epsilon_1q_2^\prime+\epsilon_2q_1^\prime-\frac{q_1^\prime q_2^\prime\chi^\parallel(\q,\omega)}{\varepsilon_0\omega^2}}
\end{align}
where $\varepsilon_0$ is the electric permittivity of free space. We further assumed the same magnetic permittivity for the two regions. The reflected mode is given by $R=T-1$.

The general case of several interfaces can be obtained from the above result by summing up the (infinite) trajectories leading to the total transmission and reflection, respectively. The explicit expressions for two graphene layers sandwiched by three dielectric media are presented in the appendix, solving the matching conditions directly. 

\subsection{Homogeneous medium}
In the following, we will discuss the special case of a homogeneous medium where the three dielectrica are the same in all three regions $\epsilon=\epsilon_i$. This allows us to treat both polarizations on the same footing. The general case is discussed in the appendix.

For two graphene layers separated by the distance $d$ and using the notation of Eq. (\ref{AnsatzA}), we obtain
\begin{align}
\label{Tequal}
T&=\frac{1}{1+\alpha_1+\alpha_2+\alpha_1\alpha_2(1-e^{-2q'd})}\;,\\
\label{Requal}
R&=-\frac{\alpha_1+\alpha_2e^{-2q'd}+\alpha_1\alpha_2(1-e^{-2q'd})}{1+\alpha_1+\alpha_2+\alpha_1\alpha_2(1-e^{-2q'd})}\;,  
\end{align}
where $\alpha_i=\chi_i^\perp/(2\varepsilon_0c^2q')$ for transverse and  $\alpha_i=-\chi_i^\parallel q'/(2\epsilon\varepsilon_0\omega^2)$ for longitudinal polarization. 

At the energy $\hbar\omega_{exp}$ with
\begin{align}
\label{RZero}
\alpha_1+\alpha_2e^{-2q'd}+\alpha_1\alpha_2(1-e^{-2q'd})=0\;,
\end{align}
there is no reflection, $R=0$, the evanescent equivalent to perfect transmission. At this energy, we have exponential amplification of the transmission amplitude
\begin{align}
\label{ExpAmplification}
T_{exp}=-\frac{\alpha_1}{\alpha_2}e^{2q'd}=-\frac{\sigma_1}{\sigma_2}e^{2q'd}\;,
\end{align}
where we introduced the (scalar) conductivities $\sigma_i$ of the two layers well defined for Fermi energies less than 1eV.\cite{Stauber10} We will also discuss the amplification related to the ratio of the two intensities at the two interfaces, see Eq. (\ref{Amplification}). The maximum is then centered around $R=-1$ where the expression divergences. This corresponds to the condition
\begin{align}
\label{ROne}
1+\alpha_2(1-e^{-2q'd})=0\;.
\end{align}  

Let us now introduce the dielectric matrix $\bar{\epsilon}(q,\omega)$ which for a homogeneous medium is given by\cite{Hwang09} 
\begin{align}
\label{Dielectic}
\bar\epsilon(q,\omega)=\left(
\begin{array}{cc}
1+\alpha_1   & \alpha_1e^{-q'd} \\
\alpha_2e^{-q'd}   & 1+\alpha_2
\end{array}
\right)
\;.
\end{align}
Genuine plasmons correspond to the zeros of the determinant  det$\bar\epsilon$,
\begin{align}
\label{DetE}
\epsilon(q,\omega)=(1+\alpha_1)(1+\alpha_2)-\alpha_1\alpha_2e^{-2q'd}\;.
\end{align}
This quantity was discussed in Ref. \onlinecite{Hwang09}. For two undamped plasmon modes (Im$\alpha_i$=0), the slopes around the two zeros are opposite and the corresponding delta functions obtained by Im$\epsilon^{-1}(q,\omega+\I0)$ thus have weights of different signs. In fact, also in the dissipative region the two modes can be related to opposite signs of the so-called ``loss'' function Im$\epsilon^{-1}(q,\omega+\I0)$, contrary to the results presented in Ref. \onlinecite{Hwang09}. 

The sign change of $\epsilon(q,\omega)$ does not violate causality since it is only the determinant of a $2\times2$-matrix response function. Apart from its zeros, it does not have any further physical interpretation. In this work, we thus prefer to discuss the eigenvalues of the $2\times2$-matrix current-current response function defined as ${\bf j}=-e^2\bar\chi {\bf A}^{ext}$ where ${\bf j}=(j_1,j_2)^T$ denotes the current in layer one and two. We obtain the following expression:
\begin{align}
\label{chi}
\bar\chi(q,\omega)=\frac{-1}{d\epsilon(q,\omega)}\left(
\begin{array}{cc}
\alpha_1(1+\alpha_2)   & \alpha_1\alpha_2e^{-q'd} \\
\alpha_1\alpha_2e^{-q'd}   & \alpha_2(1+\alpha_1)
\end{array}
\right)
\;,
\end{align}
where $d=q'e^2/(2\epsilon\varepsilon_0\omega^2)$ for longitudinal polarization and $d=-e^2/(2\varepsilon_0c^2q')$ for transverse polarization. In the following, we will discuss the sum of the two eigenvalues or, equivalently, the trace of  $\bar\chi(q,\omega)$, 
\begin{align}  
\chi(q,\omega)=\frac{-1}{d\epsilon(q,\omega)}\left(\alpha_1+\alpha_2+2\alpha_1\alpha_2\right)\;.
\end{align}
We will call $-\text{Im}\chi(q,\omega+\I0)$ the energy loss function as it provides the total plasmonic spectral weight.

%%%%%%%%%%%%%%%%%%%%%%%%%%%%%%%%%%%%%%%%%%%%%%%%%%%%%%%%%%%%%%%%%%%
% Section : Equal densities
%%%%%%%%%%%%%%%%%%%%%%%%%%%
\section{Layers with equal densities}
\begin{figure}[t]
\begin{center}
  \includegraphics[angle=0,width=0.9\linewidth]{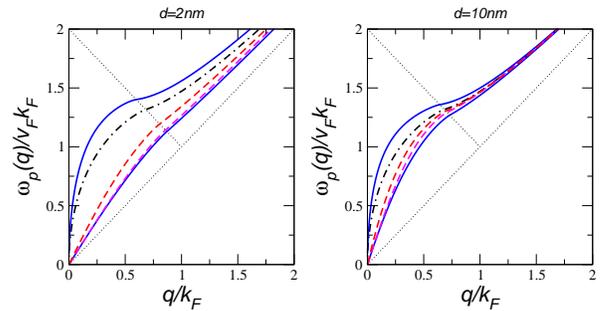}
\caption{(color online): The in-phase (upper blue curve) and out-of-phase (lower blue curve) longitudinal plasmon mode for two different layer separations $d=2,10$nm are shown together with the single layer plasmon mode obtained for $d\to\infty$ (dashed-dotted curve). Also show are the energies corresponding to $R=0$ (dashed magenta curve) and $R=-1$ (dashed red curve). Density of the two graphene layers is $n_1=n_2=10^{12}$cm$^{-2}$ with the same dielectric medium for the three regimes $\epsilon=1$ (vacuum) and we set Im$\alpha=0$.} 
  \label{figure1}
\end{center}
\end{figure}

\begin{figure}[t]
\begin{center}
  \includegraphics[angle=0,width=0.9\linewidth]{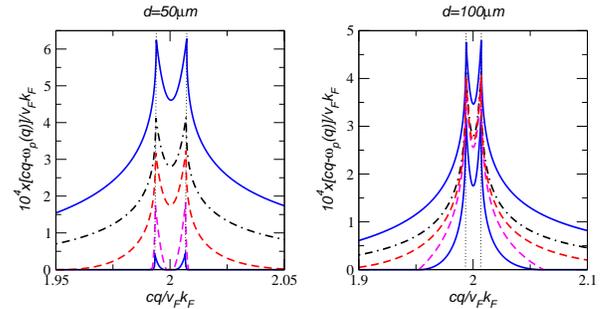}
\caption{(color online): The in-phase (upper blue curve) and out-of-phase (lower blue curve) transverse plasmon mode for two different layer separations $d=50,100\mu$m are shown together with the single layer plasmon mode obtained for $d\to\infty$ (dashed-dotted curve). Also show are the energies corresponding to $R=0$ (dashed magenta curve) and $R=-1$ (dashed red curve). Density of the two graphene layers is $n_1=n_2=10^{12}$cm$^{-2}$ with the same dielectric medium for the three regimes $\epsilon=1$ (vacuum) and we set Im$\alpha=0$.} 
  \label{figure2}
\end{center}
\end{figure}

Let us discuss some basic properties of the double-layer system in a homogeneous medium for the case of equal electronic densities, i.e., $\alpha=\alpha_1=\alpha_2$. This situation is easily realized by applying a potential difference between the two layers which results in an equal density of electrons and holes in the two layers, respectively. Since the carrier type is irrelevant for the optical properties as long as particle-hole symmetry is retained, this experimental setup is equivalent to the one with equal polar doping.

The plasmonic excitations are given by the zeros of $\epsilon(q,\omega)$, see Eq. (\ref{DetE}). For Im$\alpha$=0, this yields the following conditions satisfied by the plasmon frequency $\omega_p^\pm(q)$,
\begin{align}
1+\alpha(1\pm e^{-q'd})=0\;,
\end{align}
where the plus (minus) sign refers to the in-phase  (out-of-phase) plasmon mode. 
We will use these conditions to find resonances also in the case of so-called Landau damping (Im$\alpha\neq0$ due to particle-hole excitations), thus formally setting Im$\alpha=0$ also in the region of interband transitions ($\omega/v_F>2k_F-q$). For large layer separation $d$ and/or large wave number $q'$, the two plasmon modes thus converge to the plasmon condition for a single layer. This is expected from physical grounds since for $q'd\gg1$, the two layers are decoupled and each layer has the same plasmonic properties. In Refs. \onlinecite{Hwang09,Vazifehshenas10} the imaginary part of $\alpha$ is considered throughout and the two branches disperse when they cross the region of interband transitions. Both methods (setting Im$\alpha$=0 or not) produce similar results for weak damping, the only relevant limit to still be able to speak of a collective excitation. Still, we have found from comparison with $-\text{Im}\chi(q,\omega+\I0)$ that setting Im$\alpha=0$ also in the region of interband transitions provides a better procedure to obtain the maximum of the plasmonic spectral function. Also the energies corresponding to $R=0$ and $R=-1$ will be obtained by setting Im$\alpha=0$.

For $q'd\gg1$, Eq. (\ref{RZero}) coincides with the plasmon condition for uncoupled layers ($d\rightarrow\infty$), i.e., $1+\alpha=0$. For $q'd\ll1$, we have for the out-of-phase mode the simplified condition $1+\alpha q'd=0$. In this limit, this coincides with the condition for zero reflectivity given in Eq. (\ref{RZero}). The branch of $\omega_{exp}$ thus shifts from $\omega_p^-$ to $\omega_p^+$ as the layer separation $d$ is increased. For $d\rightarrow\infty$, $\omega_{exp}$ is located in between the two plasmonic resonances $\omega_p^\pm$ just as in the original model of Pendry with a left-handed lens.\cite{GomezSantos03} The reason why this is not the case for small layer separation is just because the two oscillators have different oscillator strengths which shifts the resonance condition.

In Fig. \ref{figure1} (longitudinal polarization) and Fig. \ref{figure2} (transverse polarization), the in-phase (upper blue curve) and out-of-phase (lower blue curve) plasmon mode for two different layer separations $d$ are shown together with the single layer plasmon mode obtained for $d\to\infty$ (dashed-dotted curve). We also show the energies corresponding to $R=0$ and $R=-1$. For both polarizations, we choose $n_1=n_2=10^{12}$cm$^{-2}$ which corresponds to $k_F\approx0.177$nm$^{-1}$ and the same dielectric medium for the three regimes with $\epsilon=1$ (vacuum).

As already discussed, in the above figures we have only considered the real part of the susceptibility and thus neglected the possibility of damping. At zero temperature, damping will occur at wave numbers $q>2k_F-\omega/v_F$ due to interband transitions. This is indicated by one of the dotted lines in the two figures. In Fig. \ref{figure1}, also the {\it limiting} dispersion for the acoustic mode $\omega=v_Fq$ is shown. Even though the ``approximate'' analytical formula of Ref. \cite{Hwang09} for longitudinal polarization states that the slope is proportional to the Fermi wave vector which could be chosen arbitrarily small, the square-root singularity of the susceptibility at $\omega=v_Fq$ forces the acoustic plasmon velocity to remain greater than $v_F$.

For transverse polarization, usually the long wavelength approximation of the susceptibility is used assuming an infinite speed of light. This yields a logarithmic divergence for $\omega/v_Fk_F=2$.\cite{Ziegler07} Including the full $q$-dependence, we see two finite maxima at $\omega/v_Fk_F=2/(1\pm v_F/c)$ when the light cone crosses the regions defined by the straight lines $\omega/v_F=2k_F\pm q$. 
\subsection{Damping and amplification of evanescent modes}

\begin{figure}[t]
\begin{center}
  \includegraphics[angle=0,width=\linewidth]{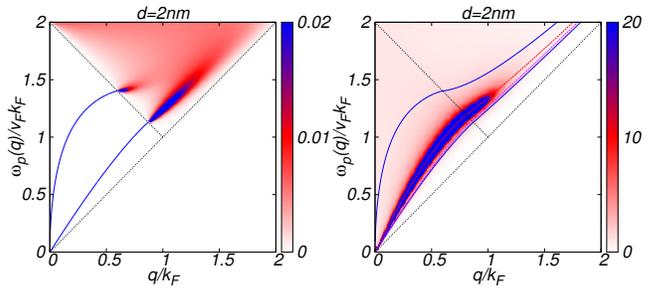}
  \caption{(color online): Left hand side: The energy loss function $-\text{Im}\chi(q,\omega+\I0)$ in units of $\hbar=v_F=1$ for longitudinal polarization. Right hand side: The amplification $\mathcal{A}$ of longitudinal polarization together with the plasmonic dispersion of the in-phase and out-of-phase plasmon mode (solid blue curves). Also the energies corresponding to $R=0$ (dashed magenta curve) and $R=-1$ (dashed red curve) are shown. The density of the two graphene layers is $n_1=n_2=10^{12}$cm$^{-2}$ with the same dielectric medium for the three regimes $\epsilon=1$ (vacuum).} 
  \label{figure3}
\end{center}
\end{figure}

\begin{figure}[t]
\begin{center}
  \includegraphics[angle=0,width=\linewidth]{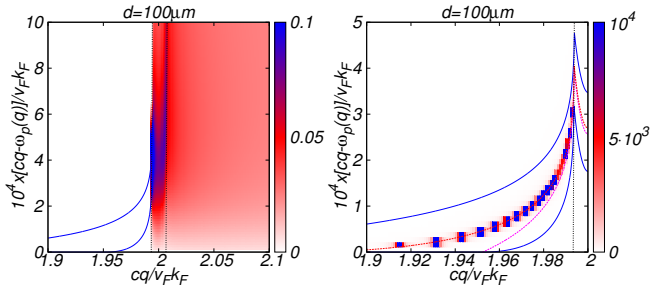}
  \caption{(color online): Left hand side: The energy loss function $-\text{Im}\chi(q,\omega+\I0)$ in units of $\hbar=v_F=1$ for transverse polarization. Right hand side: The amplification $\mathcal{A}$ for transverse polarization together with the plasmonic dispersion of the in-phase and out-of-phase plasmon mode (solid blue curves). Also the energies corresponding to $R=0$ (dashed magenta curve) and $R=-1$ (dashed red curve) are shown. The density of the two graphene layers is $n_1=n_2=10^{12}$cm$^{-2}$ with the same dielectric medium for the three regimes $\epsilon=1$ (vacuum).} 
  \label{figure4}
\end{center}
\end{figure}

To discuss the damping of the plasmonic mode, we plot the energy loss function, $-\text{Im}\chi(q,\omega+\I0)$. The results are shown on the left hand side of Fig. \ref{figure3} (longitudinal polarization) and Fig. \ref{figure4} (transverse polarization). Introducing a infinitesimal imaginary part to the frequency, $\omega\rightarrow\omega+\I0$, gives rise to delta functions defining the plasmonic frequencies in the region where there is no damping. These are shown as solid blue lines, which broaden for $q>2k_F-\omega/v_F$.

Let us now discuss the amplification of evanescent modes. For this, we introduce $\mathcal{A}$ proportional to the ratio of the field intensity at the second interface with respect to the one at the first interface:
\begin{align}
\label{Amplification}
\mathcal{A}=\left|\frac{T}{1+R}\right|^2\;.
\end{align}
The amplification diverges for $R=-1$ and the correspding energy lies in between the two plasmon modes as seen from Eq. (\ref{ROne}), but does not coincide with the exponential amplification given in Eq. (\ref{RZero}) (only in the limit $d\rightarrow\infty$, the two expressions converge). This can be seen on the right hand side of Figs. \ref{figure3} (longitudinal polarization) and \ref{figure4} (transverse polarization), where the amplification $\mathcal{A}$ is shown together with the energies corresponding to $R=0$ (dashed magenta curve) and $R=-1$ (dashed red curve). We note that even after including a small finite damping term, the amplification is strongly peaked around the red dashed curve ($R=-1$) and only the grafical cutoff makes $\mathcal{A}$ appear to be smeared out.

For large layer separation, though, the two plasmon modes coincide and also the maximum of $\mathcal{A}$ is centered around this curve as well as $\omega_{exp}$. This is not shown here in the case of equal densities, but the situation is similar to the one shown on the right hand side of Fig. \ref{figure6} and Fig. \ref{figure7}.

\subsection{Detecting transverse plasmons} 
For transverse polarization, the plasmon modes are close to the light cone.\cite{Ziegler07} We, therefore, introduced the adimensional variable $Q=cq/(v_Fk_F)$ and showed the difference $Q-\omega/v_Fk_F$ in Fig. \ref{figure2} and \ref{figure4}. The fact that they are almost pinned to the light cone makes them difficult to observe. Recently, it was proposed to use strained graphene to detect them, since strain slightly changes the dispersion relation.\cite{Pellegrino11} An alternative detection method is based on fluorescence quenching of graphene which is entirely determined by transverse plasmons in an appropriate energy window.\cite{Gomez11} Here, we want to stress that also the double-layer geometry could be used to indirectly detect these collective excitations which are linked to the two-band structure of the carriers of graphene. 

The main idea is that exponential amplification is linked to the two plasmon modes. By increasing the interlayer distance $d$, the amplification can become arbitrarily large, see Eq. (\ref{ExpAmplification}). But for large separation, the interaction between the two layers decreases and dissipative effects will eventually render the effect. 

Let us conclude with some numbers. For $n=10^{12}$cm$^{-2}$, the layer separation should be larger than $d\gapp50\mu$m, the distance where the second (out-of-phase) mode appears. For larger separation $d\gapp1$mm, the two plasmon modes almost coincide, and the exponential amplification $\omega_{\exp}$ lies almost on top of these modes. We would thus expect an optimal separation to lie around the two limiting separations, i.e., $d\approx100\sim500\mu$m, suitable to observe strong amplification.

%%%%%%%%%%%%%%%%%%%%%%%%%%%%%%%%%%%%%%%%%%%%%%%%%%%%%%%%%%%%%%%%%%%
% Section : Different densities
%%%%%%%%%%%%%%%%%%%%%%%%%%%
\section{Layers with different densities}
Let us now assume a density difference between the two layers. For longitudinal plasmons, we will investigate the case where one layer has a typical density of $n_1=10^{12}$cm$^{-2}$ and the other layer a (slightly) lower doping level with $n_2=10^{11}$cm$^{-2}$. We note that even at the neutrality point there are (damped) plasmons at finite temperature which can be treated with the analytical formulas given above for finite $k_F\rightarrow \frac{kT}{\hbar v_F}2\ln2$.\cite{Vafek06} At room temperature, this corresponds approximately to $n=10^{11}$cm$^{-2}$, but in this case also strong damping terms are present.

\begin{figure}[t]
\begin{center}
  \includegraphics[angle=0,width=\linewidth]{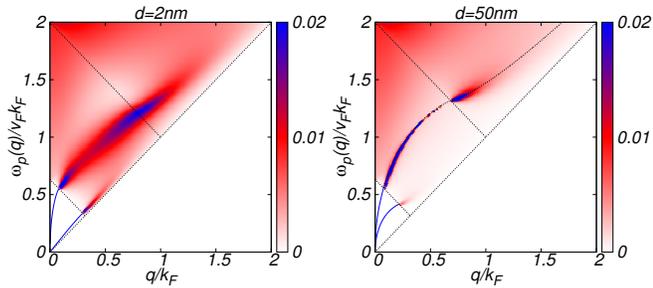}
\caption{(color online):  Left hand side: The energy loss function $-\text{Im}\chi(q,\omega+\I0)$ in units of $\hbar=v_F=1$ for longitudinal polarization and two different layer separations $d=2$nm (left) and $d=50$nm (right). Right hand side: Also shown the plasmonic dispersion of the in-phase plasmon mode as obtained without the imaginary part of $\alpha_i$ (black). The density of the two graphene layers is $n_1=10^{12}$cm$^{-2}$ and $n_2=10^{11}$cm$^{-2}$ with the same dielectric medium for the three regimes $\epsilon=1$ (vacuum).} 
  \label{figure5}
\end{center}
\end{figure}

Due to the lesser lateral confinement of the plasmon on the layer with lower density, larger layer separations are now possible until convergence of the two plasmon modes sets in. By convergence we now mean that the two plasmon modes are defined by the free conditions $1+\alpha_1=0$ and $1+\alpha_2=0$ in their respective energy window without interband transitions. In Fig. \ref{figure5}, we show the energy loss function for two different layer separations $d=2$nm and $d=50$nm. There are thus two regimes: in the first regime with $d\lapp20$nm, there are an acoustic and a plasmonic-like mode. Both modes are only undamped in the region where no interband transitions are present. When interband transitions due to the lower density layer set in, both modes are damped but the plasmonic one reaches up to the region where interband transitions due to the higher density set in. For the second regime with $d\gapp20$nm, there is a almost free plasmon with $1+\alpha_1\approx0$ for $q>2k_F^2-\omega/v_F$ limited only by interband transitions related to the higher density. Also the free plasmon $1+\alpha_1=0$ is shown as dashed black line as guide for the eye. One the other hand, the acoustic mode has become plasmonic with $1+\alpha_2\approx0$, limited by interband transitions related to the lower density.

\begin{figure}[t]
\begin{center}
  \includegraphics[angle=0,width=\linewidth]{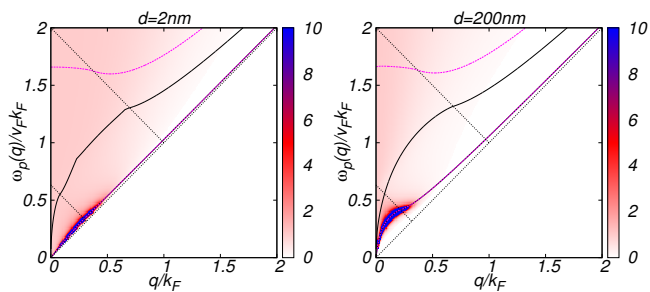}
\caption{(color online): The amplification $\mathcal{A}$ for longitudinal polarization and two different distances $d=2$nm (left) and $d=200\mu$m (right) choosing $n_2=10^{11}$cm$^{-2}$ and $n_1=10^{12}$cm$^{-2}$. The plasmonic modes (black) and the conditions for exponential amplification (magenta) with Im$\alpha_i=0$ are also shown.} 
  \label{figure6}
\end{center}
\end{figure}

The condition for exponential amplification depends on whether the electromagnetic wave first passes the high density layer or the low density layer, see Eq. \ref{RZero}. In Fig. \ref{figure6}, the amplification $\mathcal{A}$ is shown for the first case, choosing the layer densities as $n_1=10^{12}$cm$^{-2}$ and $n_2=10^{11}$cm$^{-2}$. The amplification $\mathcal{A}$ follows the out-of-phase mode and terminates at the region where interband transitions related with the lower density set in.

Also shown is the condition for exponential amplification (magenta curve for Im$\alpha_i=0$) which yields a low and a high-energy solution. The low energy solution follows the low energy plasmon mode for both regimes (left and right panel). The high energy solution starts at $\omega/(v_Fk_F^1)=1.667$ and then disperses towards higher energy roughly following the plasmonic in-phase mode. 

\begin{figure}[t]
\begin{center}
  \includegraphics[angle=0,width=\linewidth]{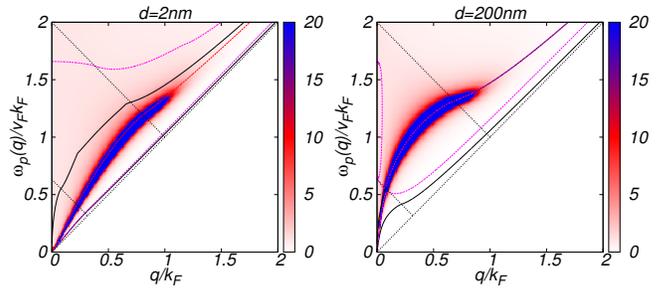}
\caption{(color online): The amplification $\mathcal{A}$ for longitudinal polarization and two different distances $d=2$nm (left) and $d=200$nm (right) choosing $n_1=10^{11}$cm$^{-2}$ and $n_2=10^{12}$cm$^{-2}$. The plasmonic modes (black) conditions for exponential amplification (magenta) with Im$\alpha_i=0$ are also shown. Additionally, on the left hand side, we present the energy corresponding to $R=-1$ (dashed red curve).} 
  \label{figure7}
\end{center}
\end{figure}

In Fig. \ref{figure7}, the the amplification $\mathcal{A}$ together with the plasmon modes (setting Im$\alpha_i=0$) is shown for the opposite layer densities, thus choosing $n_1=10^{11}$cm$^{-2}$ and $n_2=10^{12}$cm$^{-2}$.  For (very) small layer separation $d\ll20$nm, the maximum of $\mathcal{A}$ is centralized in the vicinity of the out-of-phase mode (not shown), then crosses the region between the two plasmonic modes (left hand side), until it reaches the in-phase mode for large layer separation $d\gapp20$nm (right hand side). 

The amplification now only terminates at the region defined by interband transitions related with the {\em higher} density. This can be understood from Eq. (\ref{ROne}), which only contains the physical properties of the second (higher-energy) layer. This condition is shown for Im$\alpha_2=0$ as dashed red curve on the left hand side of Fig. \ref{figure7} and is slightly shifted towards higher energies than the maximum of $\mathcal{A}$ due to the neglect of damping terms. Again we stress that $\mathcal{A}$ has a pronounced resonance close to the energy corresponding to $R=-1$ and only appears smeard out due to the chosen grafical cutoff.

The condition for exponential amplification (magenta curve for Im$\alpha_i=0$) now yields different solutions depending on the layer separation. For $d\lapp20$nm, we again find a low and a high-energy solution. For the second regime $d\gapp20$nm, the two solutions for exponential amplification split. There is a branch which only exists at small wavevectors and within an energy window reaching from  $\omega/(v_Fk_F^>)\approx1.667$ to $\omega/(v_Fk_F^<)\approx2$. The second branch starts at $\omega/(v_Fk_F^<)\approx2$ for $q=0$ and then follows the in-phase mode when it crosses the region of interband transitions coming from the lower density layer. The third branch first follows the in-phase mode for small $q$ wave vectors and then disperses towards smaller energies at the (avoided) crossing point with the second branch. We note that this crossing happens at the energy $\omega/(v_Fk_F^<)\approx1.667$. The repeated occurrence of the value $\omega/(v_Fk_F^i)\approx1.667$ is due to the fact, that at this energy the response of the $i$th graphene sheet becomes zero, which obviously affects the response of the other graphene sheet.

Let us close with a comment on transverse plasmons. For this polarization, there is generally only a small window in which undamped plasmons can exist, i.e., $1.667<\Omega<2$ with $\Omega=\omega/(v_Fk_F)$.\cite{Ziegler07} Since for exponential amplification, the two plasmons of the two layers have to exist at the same frequency, this limits the density difference to $k_F^>/k_F^<\lapp1.2$. But from Fig. \ref{figure2}, we see that the plasmon with lower energy exists in an even smaller energy window and we obtain solutions for typical densities $n=10^{12}$cm$^{-2}$ only for $k_F^>/k_F^<\leq1.02$. Nevertheless, there is a rich phenomenology also in the case of a larger density difference. We leave this analysis when detailed experimental setups are available.

%%%%%%%%%%%%%%%%%%%%%%%%%%%%%%%%%%%%%%%%%%%%%%%%%%%%%%%%%%%%%%%%%%%
% Section : Summary and Conclusions
%%%%%%%%%%%%%%%%%%%%%%%%%%%
\section{Summary and conclusions} 
We have discussed the optical properties of double layer graphene for linearly polarized evanescent modes and focused on the plasmon dispersion and (exponential) amplification of evanescent modes. We considered longitudinal as well as transverse polarization and especially the detection of exponential amplification for transverse polarized modes might at last yield evidence of transverse plasmons, uniquely related to the chirality of the electronic carriers of graphene and not found in other materials.

For systems with different densities for the two layers, we observe that the transmission crucially depends on the the order of the two layers. If the first layer is the high density layer, the amplification of the evanescent modes follows the dispersion of the low energy solution, whereas for the other case, the high energy solution is relevant. By varying the relative density of the two layers, one can switch from amplifications based on the out-of-phase mode to ones based on the in-phase mode. This might give rise to interesting applications where the modulation of an external gate can be used for an optical switcher or, more generally, for a tunable near field amplifier. 

\section{Acknowledgments} 
We thank E. H. Hwang for comments on the manuscript. This work has been supported by grants PTDC/FIS/101434/2008 and FIS2010-21883-C02-02.

\appendix
\section{General formulas for three different dielectrica}
\begin{figure}[t]
\begin{center}
  \includegraphics[angle=0,width=\linewidth]{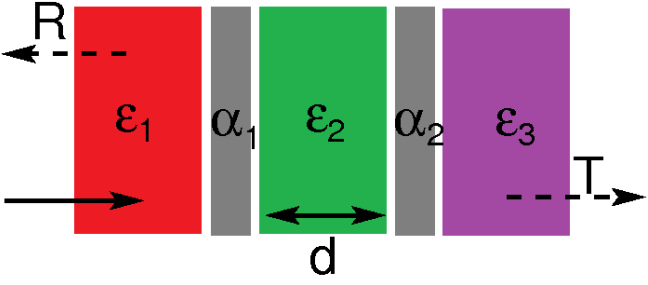}
\caption{(color online): Schematic setup of the double-layer graphene structure. The two graphene layers, characterized by $\alpha_1$ and $\alpha_2$ and separated by a distance $d$, are sandwiched by the three dielectric media characterized by $\epsilon_1$, $\epsilon_2$, and $\epsilon_3$. The income wave (normalized to one) is reflected ($R$) and transmitted ($T$).} 
  \label{figure8}
\end{center}
\end{figure}

Here, we present the general formulas of the reflectivity and transmissivity for a structure with three different dielectrica and graphene in there interfaces, i.e., there are three dielectric media separated by graphene located at $z=0$ and $z=d$. The graphene layers shall be parallel to the $xy$-plane. A schematic representation of the setup is shown in Fig. \ref{figure8}.
We will discuss the scattering problem only for the vector potential. We thus work with a gauge in which the scalar potential is set to zero. 

For transverse polarized light with say $\q=q \e_x$, only the $y$-component $A_y$ is non-zero. Since we are interested in the time-evolution of evanescent modes, we define the {\em real} transverse wave number $q_i^\prime=\sqrt{q^2-\epsilon_i(\omega/c)^2}$, where $q$ denotes the conserved wavenumber parallel to the plane and $\epsilon_i$ dielectric constant of the corresponding medium. 

\subsection{Transverse polarization}
The vector potential for a transversed polarized filed with $\q=q \e_x$ is given by
\begin{align}
\label{app:AnsatzA}
A_y(\r,z)=\sum_{q_x} e^{\I q_xx}\left\{
\begin{array}{ll}
e^{-q_1'z}+Re^{q_1'z}&,z<0\\
ae^{-q_2'z}+be^{q_2'z}&,0<z<d\\
Te^{-q_3'z}&,z>d
\end{array}
\right.\;. 
\end{align}
The vector field is continuous at the interface. From $\nabla \times{\bf B}=\mu_0 {\bf j}$, it follows that the first derivative of $A_y$ is discontinuous at the interface due to the current of the form ${\bf j}=j_i{\bf e}_y\delta(z-z_i)$ with $z_i=0,d$. Since the current is generated by the vector field, i.e., $j_i=-\chi_i^\perp A_y(z_i)$ where $\chi_i^\perp$ is the current-current response function of the graphene layer at $z=z_i$, the set of equations closes and we obtain the following expressions:
\begin{align}
a&=\frac{2q_1'(q_2'+q_3'+2\alpha_2)}{N}\\
b&=\frac{2e^{-2q_2'd}q_1'(q_2'-q_3'-2\alpha_2)}{N}\\
N&=(q_1'+q_2'+2\alpha_1)(q_3'+q_2'+2\alpha_2)\\\notag
&-(q_1'-q_2'+2\alpha_1)(q_3'-q_2'+2\alpha_2)e^{-2q_2'd}
\end{align}
where $\alpha_i=\chi_i^\perp/(2\varepsilon_0c^2)$. This leads to 
\begin{align}
T^\perp=4q_1'q_2'e^{(q_3'-q_2')d}/N\;,\;R^\perp=a+b-1\;.
\end{align}

At the energy $\hbar\omega_{exp}$ with
\begin{align}
&(q_2'-q_1'+2\alpha_1)(q_3'+q_2'+2\alpha_2)\\\notag
=&(q_1'+q_2'-2\alpha_1)(q_2'-q_3'-2\alpha_2)e^{-2q_2'd}\;,
\end{align}
there is no reflection, $R=0$. At this energy, we have exponential amplification of the transmission amplitude
\begin{align}
T_{exp}^\perp=\frac{q_2'-q_1'+2\alpha_1}{q_2'-q_3'-2\alpha_2}e^{(q_2'+q_3')d}\;.
\end{align}

The amplification $\mathcal{A}$ diverges for 
\begin{align}
&(q_2'+q_3'+2\alpha_2)=-(q_2'-q_3'-2\alpha_2)e^{-2q_2'd}\;.
\end{align}
\subsection{Longitudinal polarization}

For longitudinal polarization, the general vector field also has a component in normal direction to the interfaces:
\begin{align}
{\bf A}({\bf r},z)=\sum_\q  e^{\I \q\cdot{\bf r}}\left(A_\parallel(\q,z){\bf e}_\q+A_\perp(\q,z){\bf e}_z\right)
%=\sum_\q e^{\I \q\cdot\r}\left({\bf
%  e}_\q ae^{-q'z}+{\bf e}_z be^{-q'z}\right)
\end{align}
where with $q_i^\prime=\sqrt{q^2-\epsilon_i(\omega/c)^2}$ we make the ansatz $(i=\parallel,\perp)$
\begin{align}
\label{app:AnsatzAlong}
A_i(\q,z)=\left\{
\begin{array}{ll}
e^{-q_1'z}+R_ie^{q_1'z}&,z<0\\
a_ie^{-q_2'z}+b_ie^{q_2'z}&,0<z<d\\
T_ie^{-q_3'z}&,z>d
\end{array}
\right.\;. 
\end{align}
The components $A_\perp$ are obtained from $A_\parallel$ via the condition for a transverse field $\nabla\cdot{\bf A}=0$. In the following we will drop the index $i$ and evaluate the coefficients for $i=\parallel$.

The vector field is continuous at the interface, but the normal component of the displacement field makes a jump due to the presence of graphene and the corresponding charge density in the layer. This component is related to the vector field via the relation $D_z=\epsilon_0\epsilon\I\omega A_z$. With the continuity equation $\omega\rho-{\bf q}\cdot{\bf j}=0$ and the linear response $j=-\chi_i^\parallel A_\q$, the set of equations closes. Assuming the same notation as in the case of transverse radiation, we obtain the following expressions:
\begin{align}
a&=\frac{2q_2'\epsilon_1(q_3'\epsilon_2+q_2'\epsilon_3+2\alpha_2)}{N}\\
b&=\frac{2e^{-2q_2'd}q_2'\epsilon_1(q_3'\epsilon_2-q_2'\epsilon_3-2\alpha_2)}{N}\\
N&=(q_2'\epsilon_1+q_1'\epsilon_2+2\alpha_1)(q_2'\epsilon_3+q_3'\epsilon_2+2\alpha_2)\\\notag
&-(q_2'\epsilon_1-q_1'\epsilon_2+2\alpha_1)(q_2'\epsilon_3-q_3'\epsilon_2+2\alpha_2)e^{-2q_2'd}
\end{align}
where $\alpha_1=-\chi_1^\parallel q_1'q_2'/(2\varepsilon_0\omega^2)$ and $\alpha_2=-\chi_2^\parallel q_2'q_3'/(2\varepsilon_0\omega^2)$. This leads to 
\begin{align}
T^\parallel=4q_2'q_3'\epsilon_1\epsilon_2e^{(q_3'-q_2')d}/N\;,\;R^\parallel=a+b-1\;.
\end{align}

At the energy $\hbar\omega_{exp}$ with
\begin{align}
&(q_1'\epsilon_2-q_2'\epsilon_1+2\alpha_1)(q_2'\epsilon_3+q_3'\epsilon_2+2\alpha_2)\\\notag
=&(q_2'\epsilon_1+q_1'\epsilon_2-2\alpha_1)(q_3'\epsilon_2-q_2'\epsilon_3-2\alpha_2)e^{-2q_2'd}\;,
\end{align}
there is no reflection, $R=0$. At this energy, we have exponential amplification of the transmission amplitude
\begin{align}
T_{exp}^\parallel=\frac{q_1'\epsilon_2-q_2'\epsilon_1+2\alpha_1}{q_3'\epsilon_2-q_2'\epsilon_3-2\alpha_2}\frac{q_3'}{q_1'}e^{(q_2'+q_3')d}\;.
\end{align}

The amplification $\mathcal{A}$ diverges for 
%\begin{align}
%&(q_3'\epsilon_2+q_2'\epsilon_3+2\alpha_2)\notag\\
%=-&(q_3'\epsilon_2-q_2'\epsilon_3-2\alpha_2)e^{-2q_2'd}\;.
%\end{align}
\begin{align}
(q_3'\epsilon_2+q_2'\epsilon_3+2\alpha_2)
=-(q_3'\epsilon_2-q_2'\epsilon_3-2\alpha_2)e^{-2q_2'd}\;.
\end{align}

%%%%%%%%%%%%%%%%%%%%%%%%%%%%%%%%%%%%%%%%%%%%%%%%%%%%%%%%%%%%%%%%%%%
% Section : Bibliography
%%%%%%%%%%%%%%%%%%%%%%%%%%%

\end{document}